\documentclass[aps,preprint]{revtex4}%
\usepackage{amsfonts}
\usepackage{amsmath}
\usepackage{amssymb}
\usepackage{graphicx}%
\begin{document}
\title[Cross correlation]{Cross correlation of low-lying levels of even-even nuclei}
\author{M. H. Simbel}
\author{A. Y. Abul-Magd}
\affiliation{Faculty of Science, Zagazig University, Zagazig, Egypt}
\keywords{Cross correlation, Low-lying levels, Even-even nuclei}
\pacs{21.60.Ev, 24.60.Lz, 02.50.Tt}

\begin{abstract}
We consider the cross correlations of low-lying energy levels of nuclei
belonging to small intervals of the excitation-energy ratio $R_{4/2}$ of the
first 4$^{+}$ to 2$^{+}$ states. The mean value of the cross-correlation
coefficients, plotted as a function of $R_{4/2}$ has deep minima at values of
$R_{4/2}=2.2$ and $2.9$, which correspond to the critical dynamical symmetries
of the interacting Boson model. The distribution of the calculated
coefficients for nuclei belonging to different $R_{4/2}$ classes has different pattern.

\end{abstract}
\startpage{1}
\endpage{2}
\maketitle

\section{INTRODUCTION}

Geometrical collective models provide a suitable framework for studying low
lying levels of several even-even nuclei \cite{greiner}. In particular, the
interacting boson model (IBM) has been very successful in describing the
collective behavior algebraically \cite{iachello}. In the simplest version of
this model, known as IBM-1, an even-even nucleus with $n$ valence nucleons is
treated as a system of $n/2$ bosons interacting via two-body force. The model
has three dynamical symmetries, obtained by constructing the chains of
subgroups of the $U(6)$ group that end with the angular momentum group
$SO(3)$. The symmetries are labeled by the first subgroup appearing in the
chain which are $U(5)$, $SU(3)$, and $O(6)$ corresponding, respectively, to
vibrational, rotational and $\gamma$--unstable nuclei. There are few nuclei
close to these symmetry limits. The majority are in an earlier or later stage
of the of a phase (or shape) transition between the dynamical symmetries.
Recently, Iachello \cite{iachello1,iachello2} introduced new dynamical
symmetries, labeled as $E(5)$ and $X(5)$ at the critical point of the
$U(5)$-$O(6)$ and $U(5)$-$SU(3)$ phase transitions, respectively.

It is not \textit{a priori} clear to what extent can one distinguish the
location of prevalence of the dynamical symmetries in the chart of nuclei.
Most of the efforts are focused on comparing the level schemes of selected
nuclei with the predictions of IBM. A variety of signatures have been employed
in the identification of the symmetry properties of numerous individual nuclei
that have large numbers of low-lying levels with definite spin-parity
assignment (see, e.g. \cite{kern}). However, individual nuclei still remain
individual. The need to associate the spectral properties of one nucleus with
another calls for the definition of a model-independent parameter that enables
one to understand the evolution of the collective nuclear structure. Among the
proposed parameters are the product of the number of valence protons relative
to the nearest closed shell and the valence neutron number, $N_{p}N_{n}$, and
the ratio of the excitation energies of the first 4$^{+}$ and the first
2$^{+}$ level in each nucleus, $R_{4/2}$\ \cite{casten1}.

During the past decades, a vast amount of nuclear spectroscopic data has been
accumulated. Level schemes involving tens and sometimes hundreds of levels
with reliably known values of spin and parity are now available for hundreds
of nuclei (see Ref.~\cite{NDS}). The wealth of published spectroscopic data
allows for an extensive study of the level statistics of nuclei at low
excitation energies. Recent statistical analyses of level spacings of
low--lying states with spin and parity 2$^{+}$ \cite{ahsw} lead to more
definitive and precise statements about regularity versus chaos in this domain
than has been possible so far. These nuclei are grouped into classes that have
common collective behavior. The classes are defined in terms of the
excitation-energy ratio $R_{4/2}$. The parameter that measures the degree of
chaoticity has deep minima at $R_{4/2}=2.0$, $2.5$, and $3.3$. These minima
correspond, respectively, to the $U(5)$, $SO(6)$, and $SU(3)$ dynamical
symmetries of the IBM.

In the present paper, we add a new criterion to the existing set to identify
the collective motion in nuclei. The low lying levels of collective nuclei are
expressed as functions of the parameters of some simple model, e.g. IBM. If a
group of nuclei belong to the same symmetry group, then levels of the same
rank (e.g., the $n$th level of given spin-parity $J^{\pi}$) in these nuclei
are strongly correlated, since their excitation energies can be evaluated by
the same formula but with different values of the model parameters. In the
case of IBM, where the level energies linearly depend on the parameters, one
even expects a linear relationship between several levels of nuclei belonging
to the same symmetry group. Cross correlation is a standard method for
estimating the degree to which two series are linearly correlated. In the
present case the series consists of the energies of a given excited state of
the nuclei belonging to the class under investigation. The the
cross-correlation coefficient (CCC) determines the extent to which the
energies of the two levels are linearly related. In this work, we calculate
CCC for each pair of the low lying levels of nuclei belonging to the same
class of the $R_{4/2}$ ratio. If the nuclei of the $R_{4/2}$-class under
investigation belongs to the same symmetry group, most of the level pairs will
have their CCC's close to 1 (or -1). On the other hand, if the nuclei fall in
the domain of phase transition between the dynamical symmetries, one expects
large fluctuations of the CCC's around a small mean value. We describe the
data in Section 2 and give a brief account of the $R_{4/2}$ classification in
Section 3. Section 4 is devoted to the calculation of CCC. The conclusion of
this work is outlined in Section 5.

\section{DATA SET}

The data on low--lying levels of even--even nuclei with spins ranging from 0
to 6 are taken from the compilation by Tilley \textit{et al.}~\cite{tilley}
for mass numbers $\ 16$ $\leq A\leq20$, from that of Endt~\cite{endt} for
$20\leq A\leq44$, and from the Nuclear Data Sheets~\cite{NDS} for heavier
nuclei. We considered within each nucleus those levels of spin--parity
$J^{\pi}$, for which the spin--parity assignments of at least five consecutive
levels are unambiguous. In cases, where the spin-parity assignments were
uncertain and where the most probable value appeared in brackets, we accepted
this value. We terminated the sequence when we arrived at a level with
unassigned $J^{\pi}$, or when an ambiguous assignment involved $J^{\pi}$ among
several possibilities, as e.g. ($J^{\pi}=\left(  2^{+},4^{+}\right)  $). We
made an exception when only one such level occurred and was followed by
several unambiguously assigned levels containing at least two $J^{\pi}$
levels, provided that the ambiguous $J^{\pi}$ level is found in a similar
position in the spectrum of a neighboring nucleus. However, this situation
occurred for less than 5\% of the levels considered. In this way, we obtained
enough levels to obtain statistically significant results. In this way, we
obtained an ensemble of 4177 levels of spin--parity $0^{+},1^{+},1^{-}%
,2^{+},2^{-},3^{+},3^{-},4^{+},4^{-},5^{+},5^{-}$ and $6^{+}$ belonging to 478
nuclei. The composition of this ensemble is shown in Table 1 where the
vacancies correspond to cases with number of levels less than three. The
number of states involved to the set is large enough to justify the
statistical analysis with few exceptions consisting mainly of states that have
no counterpart in the standard collective models.

\section{$R_{4/2}$ CLASSIFICATION}

We grouped nuclei, involved in the data set described in Section 2, into
classes. Within each class the ratio
\begin{equation}
R_{4/2}=E(4_{1}^{+})/E(2_{1}^{+})
\end{equation}
of excitation energies of the first 4$^{+}$ and the first 2$^{+}$ excited
states, must lie in a fixed interval. The width of the intervals was taken to
be 0.1 when the total number of nuclei falling into the corresponding class
was 20 or more. Otherwise, the width of the interval was increased (see Fig.
1). The use of the parameter (1) as an indicator of collective dynamics is
justified both empirically and by theoretical arguments. We recall some of
these arguments in turn.

(i) Casten \textit{et al.}~\cite{casten} plotted $E(4_{1}^{+})$ versus
$E(2_{1}^{+})$ for all nuclei with $38\leq Z\leq82$ and with $2.05\leq
R_{4/2}\leq3.15$. The authors found that the data fall on a straight line.
This suggests that nuclei in this wide range of $Z$--values behave like
anharmonic vibrators with nearly constant anharmonicity. In a subsequent
paper~\cite{zamfir} it was found that a linear relation between $E(4_{1}^{+})$
and $E(2_{1}^{+})$ holds for pre--collective nuclei with $R_{4/2}<2$. Thus,
from an empirical perspective, the dynamical structure of medium--weight and
heavy nuclei can be quantified in terms of $R_{4/2}$.

(ii) Theoretical calculations based on the IBM-1 model~\cite{iachello} support
the conclusion that $R_{4/2}$ is an appropriate measure for collectivity in
nuclei. The IBM calculation of energy levels yields values of $R_{4/2}=2.00$,
$3.33$, and $2.50$ for the dynamical symmetries $U(5)$, $SU(3)$, and $O(6)$,
respectively. In this respect, a recent systematic analysis of the NNS
distributions for 2$^{+}$ levels of even--even nuclei \cite{ahsw} aimed to
determine the chaoticity parameter $f$ for nuclei at low excitation showed
that this parameter has deep minima at these values of $R_{4/2}$. Thus, we may
expect increased regularity that leads to an enhanced correlation between the
energy levels of nuclei having any of the IBM dynamical symmetries.

(iii) Recently, Iachello has introduced a new class of symmetries that applies
to nuclei undergoing a phase transition between the limiting symmetries of
IBM. In particular, the "critical symmetry" $E(5)$ \cite{iachello1} has been
suggested to describe critical points in the phase transition from spherical
to $\gamma$-unstable shapes while $X(5)$ \cite{iachello2} describes systems
lying at the critical point in the transition from spherical to axially
deformed systems. Iachello described these symmetries using a Bohr-type
geometric Hamiltonian with a flat-bottomed potential that allows for analytic
solutions. The calculation of energy levels for this Hamiltonian yields
$R_{4/2}=2.20$ and $2.91$ for the critical symmetries $E(5)$ and $X(5)$,
respectively. We expect increased fluctuation that leads to a reduce cross
correlation between the energy levels of nuclei having $R_{4/2}$\ near one of
these values.

\section{CALCULATION OF CCC}

The cross correlation coefficient is a measure of degree to which linear model
may describe the relationship between two variables. It may take any value
between -1 and +1. A positive CCC means that as the value of one variable
increases, the value of the other variable increases; as one decreases the
other decreases. A negative CCC indicates that as one variable increases, the
other decreases, and vice-versa. Taking the absolute value of the correlation
coefficient measures the strength of the relationship between the
corresponding. Thus a CCC coefficient of zero ($C_{ij}=0,$ for variable
labeled $i$ and $j$) indicates the absence of a linear relationship and CCC's
of $C_{ij}=\pm1$ indicate a perfect linear relationship. A convenient way of
summarizing a large number of correlation coefficients is to put them in a
single table, called a correlation matrix. A correlation matrix is a table of
all possible correlation coefficients between a set of variables. One would
not need to calculate all possible correlation coefficients, however, because
the correlation of any variable with itself is necessarily 1.00. Thus the
diagonals of the matrix need not be computed. In addition, CCC is
non-directional. For this reason the correlation matrix is symmetrical around
the diagonal.

In this section we analyze CCC's for energy levels of the nuclei that belong
to each of the seventeen $R_{4/2}$\ classes, which are described in the
previous section and shown in Table 1. Collective models express level
energies of low excited states in terms of formulae, which linearly depend on
one or more the parameters. For example, the rotational model divides the
levels into bands, within which the energy of a level with spin $J$ equals
$E_{J}=(1/2I)\hslash^{2}J(J+1)$ where $I$ is the effective moment of inertia.
One thus expects a linear relationship between several levels of collective
nuclei belonging the same symmetry group. As a consequence, the mean value of
the matrix elements of their correlation matrix to be close to one.

Consider an $R_{4/2}$ class in which the spin-parity assignment of $N_{i}$
nuclei allow the identification of the level with label $i$ ($=J_{r}^{\pi})$,
which denotes the spin $J$ and parity $\pi$ of the level and its rank $r$ of
excitation (i.e., 1st , 2nd, etc) within the spin-parity group. Let
$E_{i}\left(  \nu\right)  $ be the excitation energy of the level $i$ in a
nucleus, labeled by $\nu$, belonging to this class . Then, CCC is defined as%
\begin{equation}
C_{ij}=\frac{%
{\displaystyle\sum\limits_{\nu=1}^{N}}
\left[  E_{i}(\nu)-\overline{E_{i}}\right]  \left[  E_{j}(\nu)-\overline
{E_{j}}\right]  }{\sqrt{%
{\displaystyle\sum\limits_{\nu=1}^{N}}
\left[  E_{i}(\nu)-\overline{E_{i}}\right]  ^{2}}\sqrt{%
{\displaystyle\sum\limits_{\nu=1}^{N}}
\left[  E_{j}(\nu)-\overline{E_{j}}\right]  ^{2}}},
\end{equation}
where $\overline{E_{i}}$ and $\overline{E_{j}}$ are the mean values of
energies of levels $i$ and $j$ of nuclei belonging to this class,
e.g.\ $\overline{E_{i}}=(1/N_{i})\sum\limits_{\nu=1}^{N_{i}}E_{i}(\nu)$. The
sums run over the nuclei within the class and $N=\min\left(  N_{i}%
,N_{j}\right)  $. We substitute for $E_{i}(\nu)$ in Eq. (2) the values of the
excitation energies of the levels listed in Table1. Thus, for each $R_{4/2}%
$-class, we constructe an $18\times18$ correlation matrix having unit diagonal
elements and off-diagonal elements calculated by using Eq. (2). We then
calculate the mean value $\left\langle C\right\rangle $ and standard deviation
$\sigma$ of the elements of each of these matrices. The result of calculation
are shown in Fig. 1. Most of the mean values of the CCC's are close to 1,
which suggests that nuclei belonging to the same $R_{4/2}$-class indeed have
similar structure. A particularly enhanced values of $\left\langle
C\right\rangle $ are observed in the classes of nuclei belonging to the
regions of $R_{4/2}=1.95-2.05$, $R_{4/2}=2.25-2.85$ with a summit at
$R_{4/2}\approx2.5$, and $R_{4/2}=3.25-3.33$ that correspond to the $U(5)$,
$SO(6)$, and $SU(3)$ dynamical symmetries. Another maximum is observed in the
class of $R_{4/2}=1.45-1.65$ which consists mainly of nuclei with a single
closed shell. Two deep minima of $\left\langle C\right\rangle $ are observed
in the intervals of $R_{4/2}=2.05-2.15$ and $R_{4/2}=2.85-2.95$. The histogram
for $\sigma$ has maxima at these positions. These regions are nearly at the
values $R_{4/2}=2.20$ and $2.91$ corresponding the critical symmetries $E(5)$
and $X(5)$, respectively. Another broad minimum occurs at $R_{4/2}\approx1.8$,
corresponding to a transition from non-collective (with single or double
closed shells) to collective dynamics. The standard deviations are large for
values of $R_{4/2}$, which correspond to these minima of $\left\langle
C\right\rangle $. It is well known from thermodynamics that systems undergoing
phase transitions suffer from large-scale fluctuations. Therefore, there is no
wonder that the CCC has minima and its standard deviation has maxima at
locations expected for the critical points of the phase transitions between
dynamical symmetries.

Figures 2 and 3 show the probability density function $P(C_{ij})$ of CCC's for
the classes that respectively correspond to the maxima and minima of the
histogram in Fig. 1. The four distributions shown in Fig. 2 are essentially
different from those of Fig. 3. About 80 \% of the coefficients of the four
classes in Fig. 2 are greater that 0.95 and practically none is less that
0.75. In contrast, the coefficients in Fig. 3 have wide distributions that
extends deeply into the domain negative values. However, most of the
coefficients still have large positive values, implying that
positively-correlated behavior is considerably more prevalent that
negatively-correlated (anti-correlated) behavior even in the transition
between dynamical symmetries.

\section{SUMMARY AND CONCLUSION}

Since the proposal of IBM, several papers have been devoted to the
identification of nuclei belonging to the $U(5)$, $SO(6)$, and $SU(3)$
dynamical symmetries. Several stringent criteria has been proposed for this
purpose, which are mainly based on the comparison between the level schemes of
selected nuclei and the predictions of IBM. A question has been raised
concerning the definition of an empirical characteristic that allows to select
candidate nuclei for testing dynamical symmetries and serve as a "control
parameter" for the related structural phase transitions. Among the proposed
characteristics is the ratio $R_{4/2}$ of the excitation energies of the
lowest 4$^{+}$ and 2$^{+}$ levels. Some of the argument in favor of
classifying nuclei according to the $R_{4/2}$ are given in Section 3. We have
shown in a previous paper that the statistical analysis of energy levels of
nuclei belonging to fixed intervals of $R_{4/2}$ may contribute to the
selection. This analysis suggests that nuclei belonging to the $R_{4/2}$
domains expected for the three dynamical symmetries have less chaotic dynamics
than other nuclei.

In the present paper, we have carried out a systematic analysis of the
collective behavior of the low-lying states of even--even nuclei by
calculating the CCC's for pairs of levels taken from similar nuclei. As the
measure of similarity we have taken the ratio $R_{4/2}$. As seen in Figure~1,
the mean value $\left\langle C\right\rangle $ of the CCC's is indeed dependent
on $R_{4/2}$. It is largest at $R_{4/2}=2.0$, $2.5$, and $3.3$. These maxima
correspond, respectively, to the $U(5)$, $SO(6)$, and $SU(3)$ dynamical
symmetries of the IBM as well as a plateau at CCC \symbol{126}0.95 around
$R_{4/2}=1.5$. It has deep minima at $R_{4/2}=2.10\pm0.05$ and $2.85\pm1.0$.
These minima correspond, respectively, to the $E(5)$ and $X(5)$ critical
symmetries that have been recently introduce to describe phase transitional
behavior. The standard deviations\ of the CCC's are much larger at these
minima that at other values of $R_{4/2}$, adding further evidence that the
minima are spotting phase transitions. Figures 2 and 3 show the distribution
CCC's for individual classes of $R_{4/2}$, indicating that the behavior is
different at the maxima and minima of $\left\langle C\right\rangle $.

In conclusion, we have established a correspondence between CCC of low-lying
levels of nuclei in narrow ranges of $R_{4/2}$ and the collective behavior of
nuclei as expressed by IBM. This finding may be added to the diverse attempt
to identify nuclei that could be located at the critical points of
shape/structural transitional regions, e.g. in \cite{garcia}. We note that the
analysis of the statistics of energy levels alone is not sufficient for
testing the collective behavior . Statistical analysis of other quantities
such as transition probabilities are required to do so. For example, recent
IBM calculation \cite{brentano} show that the E0-transition strengths are
particularly large in spherical-deformed transition regions. The analysis of
CCC's for transition strengths of nuclei belonging to classes of $R_{4/2}$ is
in progress.

\bigskip\pagebreak

Table 1. Number of levels considered for nuclei belonging to different classes
of the energy ratio $R_{4/2}.$ The notation $J_{r}^{\pi}$ is used to label a
level with spin $J$, parity $\pi$ and rank $r$.

$%
\begin{array}
[c]{ccccccccccccccccccc}%
R_{4/2}\backslash\text{ Level} & 0_{1}^{+} & 0_{2}^{+} & 1_{1}^{+} & 1_{1}^{-}
& 2_{1}^{+} & 2_{2}^{+} & 2_{3}^{+} & 2_{1}^{-} & 3_{1}^{+} & 3_{2}^{+} &
3_{1}^{-} & 3_{2}^{-} & 4_{1}^{+} & 4_{2}^{+} & 4_{1}^{-} & 5_{1}^{+} &
5_{1}^{-} & 6_{1}^{+}\\
&  &  &  &  &  &  &  &  &  &  &  &  &  &  &  &  &  & \\
1.058-1.449 & 15 & 10 & 6 & 4 & 27 & 17 & 14 & 4 & 9 & 4 & 20 & 7 & 27 & 13 &
11 & 7 & 12 & 22\\
1.45-1.649 & 17 & 11 & 3 & 6 & 23 & 16 & 13 & 3 & 3 & - & 10 & 7 & 23 & 11 &
5 & - & 12 & 15\\
1.65-1.749 & 10 & 4 & 3 & 3 & 19 & 14 & 8 & - & 6 & - & 9 & - & 19 & 8 & 3 &
3 & 5 & 11\\
1.75-1.849 & 12 & 8 & - & 5 & 24 & 18 & 10 & - & 7 & - & 14 & 5 & 24 & 13 &
4 & 3 & 10 & 20\\
1.85-1.949 & 15 & 8 & - & - & 20 & 16 & 12 & - & 7 & - & 15 & 5 & 20 & 15 &
3 & - & 14 & 14\\
1.95-2.049 & 13 & 4 & - & - & 20 & 16 & 12 & - & 12 & 3 & 12 & 7 & 20 & 16 &
6 & 3 & 10 & 13\\
2.05-2.149 & 18 & 13 & 6 & 5 & 28 & 20 & 19 & 3 & 13 & 4 & 15 & 4 & 28 & 19 &
4 & 4 & 14 & 25\\
2.15-2.249 & 14 & 9 & 3 & 6 & 26 & 16 & 13 & 5 & 8 & 5 & 17 & 7 & 26 & 13 &
6 & 6 & 13 & 18\\
2.25-2.349 & 30 & 23 & 5 & 11 & 47 & 35 & 26 & 5 & 21 & 4 & 30 & 11 & 47 &
28 & 12 & 12 & 25 & 38\\
2.35-2.449 & 20 & 14 & 4 & 5 & 30 & 23 & 19 & 4 & 18 & 4 & 21 & 4 & 30 & 18 &
7 & 13 & 15 & 21\\
2.45-2.549 & 30 & 18 & 10 & 8 & 36 & 30 & 26 & 3 & 28 & 9 & 20 & 8 & 36 & 25 &
- & 14 & 18 & 26\\
2.55-2.649 & 16 & 10 & 6 & 4 & 25 & 20 & 14 & - & 17 & 5 & 12 & 7 & 25 & 17 &
6 & 6 & 9 & 18\\
2.65-2.749 & 13 & 4 & - & 3 & 21 & 17 & 14 & - & 11 & - & 9 & - & 21 & 14 &
3 & 9 & 9 & 15\\
2.75-2.949 & 14 & 5 & - & 5 & 28 & 21 & 13 & - & 12 & - & 13 & - & 18 & 17 &
7 & 9 & 12 & 23\\
2.95-3.149 & 15 & 7 & - & 6 & 21 & 16 & 13 & 5 & 11 & 3 & 10 & 4 & 21 & 9 &
7 & 7 & 8 & 19\\
3.15-3.249 & 14 & 10 & 3 & 8 & 25 & 16 & 14 & 8 & 14 & 4 & 15 & 4 & 25 & 14 &
13 & 10 & 12 & 22\\
3.25-3.333 & 42 & 26 & 9 & 33 & 58 & 46 & 37 & 34 & 36 & 21 & 43 & 28 & 58 &
39 & 35 & 25 & 36 & 53
\end{array}
$

\bigskip\pagebreak

\bigskip

{\huge Figure Captions }

Figure 1. Mean value and standard deviation of the correlation coefficients
for all levels considered for different classes of the $R_{4/2}$ ratio.

Figure 2. Distribution of level CCC for nuclei belonging to classes of
$R_{4/2}$that involve pre-collective, vibrational, $\gamma$-unstable and
rotational nuclei.

Figure 3. Distribution of level CCC for nuclei belonging to classes of
$R_{4/2}$that involve nuclei undergoing shape/structural phase transitions.

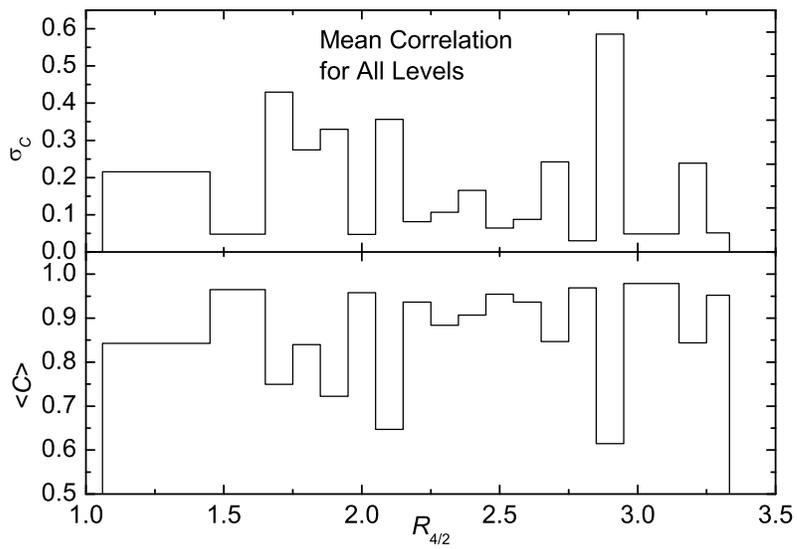
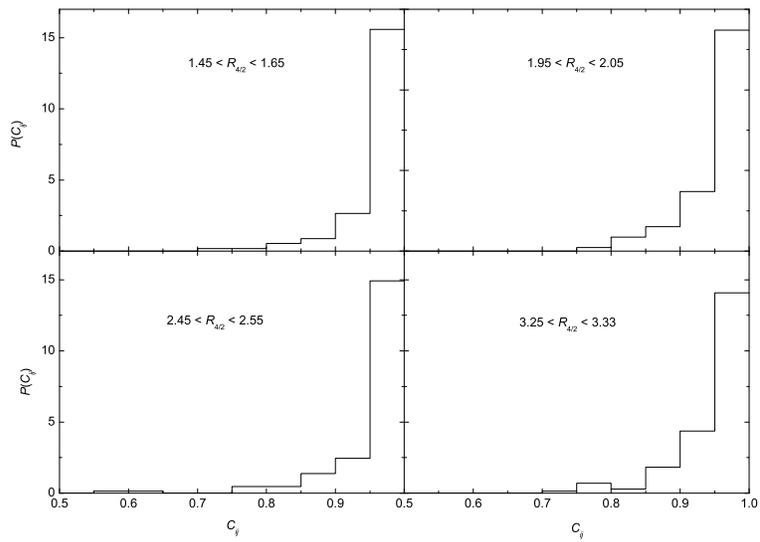
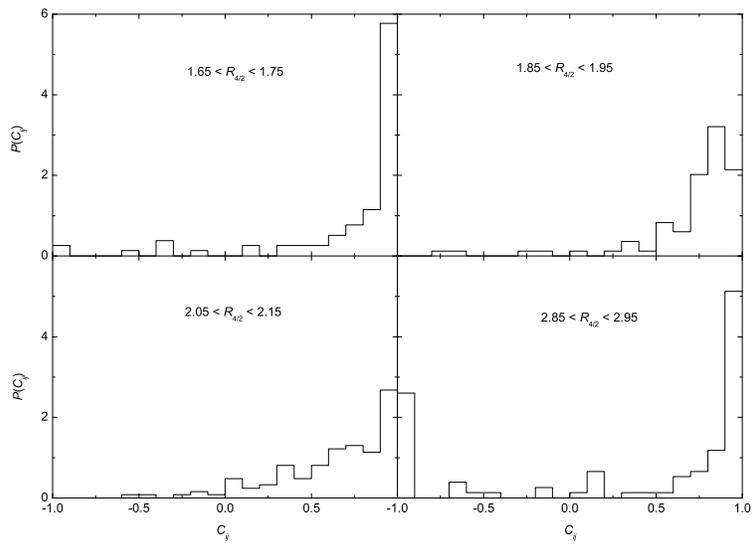
\end{document}